\documentstyle[twoside,psfig]{article}
\newcommand{\be}{\begin{equation}}
\newcommand{\ee}{\end{equation}}

\begin{document}

\pagestyle{myheadings}
\markboth{\hfill{\protect\footnotesize\rm R. Moessner \& R.
Brandenberger}\hfill}
{\hfill{\protect\footnotesize\rm Early Objects from Cosmic Strings}\hfill}

\begin{titlepage}
\end{titlepage}

\noindent {\bf ASTROPHYSICS REPORTS (Special Issue)} \hfill No. 2, {\it Page
00--00}

\noindent {\small Publ. Beijing Astronomical Observatory} \hfill January 1997

\noindent $\hrulefill$

\medskip
\bigskip
\begin{center}
\begin{LARGE}
EARLY OBJECTS IN THE COSMIC STRING THEORY WITH HOT DARK MATTER
\end{LARGE}

\bigskip
{\large Richhild Moessner$^1$ and Robert Brandenberger$^2$}

\medskip
\bigskip
{\footnotesize  $^1$Max Planck Institut f\"ur Astrophysik, 85740 Garching,
Germany

$^2$ Brown University, Department of Physics, Providence, RI
02912, USA}

\bigskip
{\small{\bf ABSTRACT}}

\end{center}

\begin{small}
\begin{quotation}
We study the accretion of hot dark matter onto moving cosmic string
loops, using an adaptation of the Zeldovich approximation to HDM.
We show that a large number of nonlinear objects of mass greater
than $10^{12}M_\odot$, which could be the hosts of high redshift
quasars, are formed by a redshift of $z=4$.
\end{quotation}

{\bf Key words}~~cosmic strings, cosmology: theory, large-scale
structure of Universe

\end{small}

\section{Introduction}
Topological defect theories provide an alternative to inflation for
explaining the origin of the primordial fluctuations which have
grown into present-day structures through gravitational instability.
Cosmic strings are one-dimensional topological defects possibly formed
in a phase transition in the early Universe.

The observed presence of massive nonlinear structures at high
redshifts of 3 to 4 has provided stringent constraints on models of
structure formation containing hot dark matter (HDM), a candidate for which
is a massive neutrino.
This is due to
the large thermal velocities of the HDM particles,  which prevent the
growth of perturbations on scales smaller than the free-streaming
length. For inflationary models with adiabatic density fluctuations,
recent data on the abundance of damped Lyman alpha
absorption systems (DLAS)$^{\cite{DLAS}}$ and on the quasar
abundance$^{\cite{QSO}}$
has restricted the fraction of HDM to less than about $30\%$ of
the total dark matter present$^{\cite{MaB}}$.
The scenario of structure formation with cosmic strings is, however, viable
even if the dark matter in the universe is hot, because cosmic strings,
which provide the seeds for the density perturbations,
survive the neutrino free-streaming$^{\cite{VS 83,BKST}}$.
In this article we present work on  the
constraints imposed on the cosmic string theory by the abundance of high
redshift quasars$^{\cite{MB}}$.

Searches for high redshift quasars have been going on for some
time$^{\cite{QSO}}$.  The quasar luminosity function is observed to rise
sharply as a function of redshift $z$ until $z \simeq 2.5$.
According to recent results from the Palomar grism survey by Schmidt et al.
(1995)$^{\cite{QSO}}$, it peaks in the redshift interval $z \, \epsilon$ [1.7,
2.7] and declines at higher redshifts.
Quasars (QSO) are extremely luminous, and it is generally assumed that they are
powered by accretion onto black holes. It is possible to estimate the mass
of the host galaxy of the quasar as a function of its luminosity,
assuming that the quasar luminosity
corresponds to the Eddington luminosity of the black hole. For a quasar
of absolute blue magnitude $M_B= -26$ and lifetime of
$t_Q=10^8 yrs$, the host galaxy mass can be
estimated as$^{\cite{PS 94}}$
\be
M_G=c_1 10^{12} M_\odot \, ,
\ee
where $c_1$ is a constant which contains the uncertainties in relating
blue magnitude to bolometric magnitude of quasars, in the baryon
fraction of the Universe and in the fraction of baryons in the host
galaxy able to form the compact central object (taken to be 10$^{-2}$).
The best estimate for $c_1$ is about 1.
Models of structure formation have to pass the test of producing
enough early objects of sufficiently large mass to host the observed
quasars.

Besides the quasars themselves, absorption lines in their spectra
due to intervening matter can be used to quantify the amount of
matter in nonlinear structures at high redshifts.
Based on the number density of absorption lines per
frequency interval and on the column density calculated from
individual absorption lines, the fraction of $\Omega$ in bound neutral gas
(denoted by $\Omega_g$) can be estimated. For high-column density
systems, the damped Ly-$\alpha$ systems (DLAS), recent observational
results are that
\be
\Omega_g (z) > 10^{-3}
\ee
for $z \, \epsilon$ [1,3]. In the most recent results by
Storrie-Lombardi et al. (1996)$^{\cite{DLAS}}$ there is evidence for a
flattening of
$\Omega_g$ at $z \sim 2$ and a possible turnover at $z \sim 3$ .
The corresponding value for $\Omega$ in bound matter
is larger by a factor of $f^{-1}_b$, where $f_b$ is the
fraction of bound matter which is baryonic, which is about $10\%$
in a flat universe.

In the next section, we give a brief review of the cosmic string scenario of
structure formation. Then we study
the accretion of hot dark matter onto cosmic string loops, which
seed large amplitude local density contrasts already at early times.
We use the
results to compute the number density of high redshift objects as a
function of a parameter $\nu$
which determines the number density of loops in the scaling solution
(see Section 2).  We demonstrate that for realistic values of $\nu$, the
number of massive nonlinear objects at redshifts $\le 4$ satisfies
the recent observational constraint of quasar abundances (see Figure~1).  We
also comment on the implications of (2).
We consider a spatially flat Universe containing HDM and baryons.
Units where $c=1$, a Hubble constant of $H=50 \,h_{50} {\rm km s^{-1}
Mpc^{-1}}$ and a redshift at equal matter and radiation of
$z_{eq}=5750 \,\Omega h_{50}^{2}$ are used.

\section{Cosmic Strings and Structure Formation}

Cosmic strings$^{\cite{CSrev}}$ are one-dimensional topological defects which
might have been formed in a phase transition in the very early
Universe. They are produced for the same
reasons as ordinary defects in condensed matter systems, such as vortex lines
in superfluid helium, but their subsequent dynamics is governed by
relativistic equations. In the simplest model, the symmetry breaking
occuring in the phase transition
is achieved via a two-component scalar Higgs field
$\phi=(\phi_{1},\phi_{2})$ with symmetry-breaking potential
$V(\phi)=\frac{1}{4} \lambda (\phi^{2}-\eta^{2})^{2}$, where
$\eta$ is the energy scale of the phase transition, and
$\lambda$ is the coupling constant. Topological considerations require
cosmic strings to have no ends, they are either formed as infinitely
long strings or as  closed loops. These strings are very thin lines of
trapped energy, the energy
being the false vacuum energy $V(\phi =0)$ which is trapped inside the defects
after the rest of the Universe has evolved to the true ground state
$|<\phi>|=\eta$ with $V(\phi) =0$. The mass per unit length of the
strings is given by the symmetry breaking scale $\eta$ , $\mu \sim
\eta^{2}$ , and it is the only free parameter of the model (in
practice uncertainties due to the  complicated evolution of cosmic
strings after the phase transition introduce extra phenomenological
parameters, which are not fundamental, however). The
energy in the defects acts gravitationally on surrounding matter and
radiation, and thus provides the origin for the density perturbations,
which evolve into today's large-scale structure.

Most important for structure formation are the infinite strings. They are
straight over distances increasing with time of one horizon distance $H^{-1}(t)
\sim t$, and form an approximate random walk on distances larger than that. In
this way they can cause structures
 to be formed on the large scales observed today. The model
makes predictions consistent with large-scale structure
observations$^{\cite{TV86,PBS}}$ and
CMB anisotropy measurements$^{\cite{CMB}}$. An intriguing fact is that the
normalization of the free parameter $\mu$ of the model to these
observations,
which gives $G \mu \approx 10^{-6}$, requires the phase transition giving rise
to the defects to take place at a scale of $\eta \sim 10^{16} GeV$ .
This is just the scale of grand unification where the three coupling
constants of the strong, weak and electromagnetic interactions meet,
suggesting
the appearance of new physics.

The network of cosmic strings is formed in a phase transition
in the early universe about $10^{-35}$ sec after the Big Bang, long
before the times relevant for structure formation. Therefore the
string network has to be evolved over many orders of magnitude in
time until $t_{eq}$ (and subsequently). This is very complicated numerically,
and would be impossible to handle analytically, if it were not
for the fact that the
network of strings quickly evolves towards a scaling solution$^{\cite{CSrev}}$
where the
energy density in long strings remains a constant fraction of the total
background energy density. This is achieved by intercommutations and
self-intersections
of the strings leading to the production of small loops, which then
decay by emitting gravitational radiation. In this way, some of the energy
input into the string network coming from the stretching of the strings due to
the expansion of the universe is transferred to the background.
According to the scaling solution, the string network looks
the same at all times when all distances are scaled by the Hubble
radius $H^{-1}(t)$, and this allows the extrapolation over such long
time intervals.
The scaling solution can be pictured as having a fixed number $M$
of long strings per Hubble volume at any given time, and a distribution of
small loops.

Some of the parameters characterizing the string network besides the
fundamental one $\mu$ are the long string velocities $v_{s}$ and
initial loop velocities $v_i$, the average number $M$ of
strings per Hubble volume, the amount of small-scale
structure on the long strings quantified by $(\mu -T)$, where $T$ is
the tension of the string, and the constants  $\alpha$, $\beta$ and
$\nu$ appearing in the next three equations. Small-scale
structure is produced mainly when string segments self-interect and
split off loops, a process which is characterized by the loop
production rate and the size of the produced loops.
According to the scaling solution, loops are formed with a radius
which is a constant fraction of the horizon,
\begin{equation}
R_f (t) = \alpha t \, ,
\end{equation}
and length
\begin{equation}
l = \beta R \, .
\end{equation}
The  number of loops per unit physical volume present at time
$t$ with lengths in the interval between $l$ and $l + dl$ is
\begin{equation}
n (l,t)dt = \nu l^{-2} t^{-2} dt\, .
\label{nlt}
\end{equation}
Since loops decay by emitting gravitational
radiation, there is a lower cutoff value of $l$ for the
distribution (5) given by $l_{min} \sim G \mu t$,
$G$ being Newton's constant.  Below $l_{min}$, the distribution
$n(l,t)$ becomes constant.

Numerical simulations$^{\cite{CSsim}}$ indicate that $\alpha \le 10^{-2}$,
$\beta \simeq 10$, and $M \sim 10$.  From these
values it follows that -- unless $\nu$ is extremely large -- most of the
mass of the string network resides in long strings (where long strings
are defined operationally as strings which are not loops with radius
smaller than the Hubble radius).

Long strings accrete matter in the form of wakes
behind them as they move through space$^{\cite{SV,TV86,PBS}}$. Spacetime around
a long straight cosmic string can be pictured as locally flat, but with  a
deficit angle$^{\cite{AV81}}$ of $8\pi G \mu$ . Therefore a string moving
relativistically with  velocity $v_{s}$ imparts velocity perturbations to
surrounding matter towards the
plane swept out by the string, of magnitude
\begin{equation}
u=4 \pi G \mu \gamma_{s} v_{s}
\end{equation}
creating overdensities in the form of a wake
behind the string.
If small-scale structure is present on the string, there is in addition a
Newtonian force towards the string, proportional to $G(\mu-T)$, the
strings move more slowly and accrete matter rather in the form of
filaments than wakes.

For HDM, the first  nonlinearities about wakes resulting from strings
without any small-scale structure
form only at late times, at a redshift of$^{\cite{PBS}}$ about 1 for
$ G\mu \simeq 10^{-6}$, which is obtained when normalizing the
model to COBE. Before this redshift, no
nonlinearities form as a consequence of accretion onto a single
uniform wake.
Thus, in the cosmic string and hot dark matter theory, a different
mechanism is required in order to explain the origin of high redshift
objects.  Possible mechanisms related to wakes are early structure
formation at the crossing sites of different wakes$^{\cite{HMM}}$,
small-scale structure of the strings giving rise to
wakes$^{\cite{Vollick,AB95}}$, and inhomogeneities inside of
wakes$^{\cite{Rees}}$.  In
this article, however, we will explore a different mechanism, namely the
accretion of hot dark matter onto loops.

In earlier work$^{\cite{BKST}}$, the accretion of hot dark matter onto static
cosmic string loops was studied.  It was found that in spite of free
streaming, the nonlinear structure seeded by a point mass grows from
inside out, and that the first nonlinearities form early on (accretion
onto string filaments proceeds similarly$^{\cite{AB95}}$). Since
loop accretion leads to nonlinear structures at high redshift, we will
now investigate this mechanism in detail to see whether enough high
redshift massive objects to statisfy the QSO constraints and (2) form.

\section{Nonlinear mass accreted by a single static loop}

We will use the modified Zeldovich approximation$^{\cite{Zeld,PBS}}$
to study the accretion of hot dark matter onto moving string
loops.  The Zeldovich approximation is a first order Lagrangian
perturbation theory technique in which the time evolution of the
comoving displacement $\psi$ of a dark matter particle from the
location of the seed perturbation is studied.
The physical distance of a dark matter particle from the center of the
cosmic string loop is written as
$$
h (q, t) = a (t) [q - \psi (q,t)] \, .
$$
For a point-like seed mass of
magnitude $m$ (the string loop in our case) located at the comoving position
$\underline{q}^\prime =
0$, the equation for $\psi$ is
\be
\left( \frac{\partial^{2}}{\partial t^{2}} + 2 \frac{\dot a}{a}
\frac{\partial}{\partial t} - 4 \pi G \rho(t) \right) \psi (q,t) =
{Gm\over{a^3 q^2}} \, .
\label{ZAeq}
\ee

This equation describes how as a consequence of the seed mass, the
motion of the dark matter particles away from the seed (driven by the
expansion of the Universe) is gradually slowed down.  If the seed
perturbation is created at time $t_i$ and the dark matter is cold,
then the appropriate initial conditions for $\psi$ are
\be
\psi (q,t_i) = \dot \psi (q,t_i) = 0 \, .
\ee

As formulated above, the Zel'dovich approximation only works for cold
dark matter, particles with negligible thermal velocities.
For HDM, free-streaming prevents the growth of perturbations on
scales $q$ below the free-streaming length $\lambda_J (t)$, which is the mean
comoving distance traveled
by neutrinos in one expansion time
$$
\lambda_J (t) = v (t) z (t) t \, ,
$$
where $v(t) \sim z (t)$ is the hot dark matter velocity. The free-
streaming length decreases with time (in comoving coordinates) as
$t^{-1/3}$, so that a scale $q$ which is initially affected by
free-streaming, $q<\lambda_J (t_i) $
becomes equal to $\lambda_J $ at some later time $t_s(q)$ , and then
remains above it and unaffected by free-streaming at later times.

A correct treatment of the effect of free-streaming requires the solution
of the collisionless Boltzmann equation for the neutrino phase-space
density$^{\cite{BKST}}$. But
a simple modification of the Zeldovich approximation for CDM gives the
same results as the full treatment$^{\cite{PBS}}$. For scales
$q<\lambda_J (t_i)$, Eq.~\ref{ZAeq} is only integrated from the time
$t_s(q)>t_i$ onwards instead
of from $t_i$ , $i.e.$ growth only starts at $t_s(q)$ , and the
initial conditions are modified to
$\psi(q,t_s(q))=0$ and $\dot \psi(q,t_s(q))=0$ . For scales where $q>\lambda_J
(t_i)$
initially the CDM formalism applies.

We can now define the mass that has gone nonlinear about a seed
perturbation as the rest mass inside of the shell which is ``turning
around", i.e. for which
$$
\dot h (q, t) = 0 \, .
$$
In the case of CDM, this yields a nonlinear mass of
\be
M_{CDM} (t) = {2\over 5} m \left( {t\over t_i}\right)^{2/3} \, ,
\label{MCDM}
\ee
and for loops where the accretion is affected by free-streaming,
\be
M_{HDM} (t) = {8 \over 125} \, {m^3\over{M_{eq}^2}} \left({t\over
t_{eq} } \right)^2
\label{MHDM}
\ee
with $M_{eq} = 2v_{eq}^3 t_{eq} /(9G)$ , $v_{eq}$ is the HDM-particle
velocity at $t_{eq}$. From Eq.~\ref{MCDM} and ~\ref{MHDM} we
can determine, for any time $t$, a mass $M^\prime(t)$ such that for all
masses $M>M^\prime(t)$ accreted by any loop so far, accretion has not been
affected by free-streaming, and Eq.~\ref{MCDM} is applicable. For
$M<M^\prime(t)$, expression~\ref{MHDM} is valid.

\section{Number density of quasar host galaxies}

We can now compute the number density $n_G(>M_G,t)$ in
nonlinear objects heavier than $M_G$  at high redshift $z(t)$ in the cosmic
string and hot dark matter model (in an analogous way  the fraction
$\Omega_{nl}$ of the critical density in such objects can be
calculated). By considering only the accretion
onto string loops we will be underestimating these quantities.

It can be shown$^{\cite{BS89}}$ that in an HDM model string loops accrete
matter independently, at least before the large-scale structure in
wakes turns
nonlinear at a redshift of  about 1, i.e., later
than the times of interest in this paper.  Hence, the number density
$n(l,t)$ of loops of length $l$ given by~(\ref{nlt}) can be combined with the
mass $M(l)$ accreted by an individual loop  to give the mass function
$n(M,t)$.  Here, $n(M,t) dM$ is
the number density of objects with mass in the interval between $M$
and $M+dM$ at time $t$.  This in turn determines the comoving density in
objects of mass $M>M_G$. Since the functional form of $M(m)$ and hence
$M(l)$ changes at
$M=M^\prime$, the functional forms of the mass function $n(M)$ will be
different above and below $M^\prime$.  Thus
\be
n_G(>M_G,t)=z^{-3}(t)  \left[ \int\limits^{M^\prime (t)}_{M_G} d M n_H (M)  +
\int\limits^{M_2 (t)}_{M^\prime (t)} d M n_C (M)  \right]  \, .
\label{nG}
\ee
Similarly, the fraction of the critical density in objects of mass greater
than $M_1$, $\Omega_{nl} (t) \, $ is
\be
\Omega_{nl} (t) =  \left[ \int\limits^{M^\prime (t)}_{M_1} d M n_H (M) M +
\int\limits^{M_2 (t)}_{M^\prime (t)} d M n_C (M) M \right] 6 \pi G t^2
\, .
\label{Onl}
\ee
Note the upper mass cutoff $M_2(t)$ introduced. It is imposed by our
approximation of loops as point masses. For masses greater than
$M_2(t)$, this approximation breaks down, the responsible loops would
have too large a radius to be able to accrete matter effectively.
$M_2(t)$ can be estimated by demanding that the mass $M(t)$ accreted
onto a loop exceed the mass
in a sphere of radius equal to the loop radius at $t_i$ . The
integral in Eq.~\ref{nG} is dominated at $M^\prime (t)$ , and can
be approximated as
\be
n_G(>M_G,t) \sim n_C (M^\prime , t) M^{\prime} z^{-3}(t) \,  \simeq
{1\over 5} \nu (\alpha \beta)^2 {\mu^3\over{M^\prime (t)^3}} z^{-3}(t) \, .
\label{nGapprox}
\ee

In Figure~1, the comoving number density of objects massive enough to
be able to host quasars, $n_G(>M_G,t)$, is plotted for  $G
\mu=10^{-6}$,  $v_{eq} = 0.1$, and the values of the parameters
$\alpha = 10^{-2}$ and $\beta=10$. The upper curve is for
static loops, the lower one for initial loop velocities of $v_i=0.25$.
The effect of loop motion is to decrease the density of these massive
objects, as discussed in the next paragraph.
We have to compare our results for $n_G(>M_G,t)$ with the number
density of host galaxies, $n_{G}^{{\rm obs}}$, inferred from the
observed quasar abundance $n_Q$. Since we assumed a finite quasar lifetime
$t_Q$, the number of host galaxies $n_{G}^{{\rm obs}}$ is
larger than $n_Q$ by a factor of $t_H/t_Q$, where $t_H$ is the Hubble time
at time $t$,
\be
n_{G}^{{\rm obs}}=\frac{t_H}{t_Q} n_Q \, .
\ee
This is also plotted in Fig.~1 .

The formalism presented above can be adapted to moving instead of
static loops$^{\cite{MB,EB}}$. Loop motion  changes  the
geometry of the accreted object, it becomes more elongated and the
transverse scale going nonlinear becomes smaller than for loops at
rest. Because of this, free-streaming of the neutrinos can hinder
the growth of perturbations for a longer time. The smallest mass
accreted at time $t$ which has been unaffected by free-streaming,
$M^{\prime}(t)$, is increased compared to the case of static loops,
thereby decreasing $n_G(t)$ according to Eq.~\ref{nGapprox}.

The result for the fraction of nonlinear mass accreted by moving loops is
\be
\Omega_{nl} (t) \sim 10^{-2} \nu \alpha_{-2} (G \mu)^2_6 v^{-1}_i z (t)^{-2}
h_{50}^{4}
\,,
\label{Onlres}
\ee
where $(G \mu)_6$ is defined as $(G \mu)/10^{-6}$.

In Fig.~1 we have plotted the curves up to a redshift of $5$. There
is another issue to consider, however. Our expresions for $n_G$ are
only valid as long as $M^\prime < M_2(t)$, due to
our approximation of loops as point sources. For  loops with an initial
velocity of  $v_i = 0.25$, this
condition is satisfied only up to a largest redshift of
\be
z_{max} = {6\over 5} \beta G \mu z_{eq} v_i^{-1} v^{-1}_{eq} \, ,
\label{zmax}
\ee
which is equal to
\be
z_{max}= 3 h^2_{50}
\ee
for $v_i = 0.25$ and $v_{eq}=0.1$ .
Beyond that redshift, the values of $n_G(t)$ and $\Omega_{nl}$ are suppressed
beyond the
results plotted in Fig.~1 and Eq.~\ref{Onlres} since only the tail of the loop
ensemble with
velocities
smaller than the mean velocity $v_i = 0.25$ manage to accrete a
substantial amount of mass. It is interesting to note that, as
mentioned in the introduction,  Storrie-Lombardi et al. (1996)$^{\cite{DLAS}}$
find a
flattening of $\Omega_g$ at $z \sim 2$ and a possible turnover at $z \sim 3$.

\section{Discussion}
We have studied the accretion of hot dark matter onto moving cosmic
string loops and made use of the results to study early structure
formation in the cosmic string plus HDM model.
The loop accretion mechanism is able to generate nonlinear objects
which could serve as the hosts of high redshift quasars much earlier
than the time cosmic string wakes start becoming nonlinear (which for $G \mu =
10^{-6}$ and $v_s = 1/2$ occurs at a redshift of about 1).

The fraction $\Omega_{nl} (z)$ of the total mass accreted into nonlinear
objects by string loops unfortunately depends very sensitively on
$\alpha$ and $\nu$.  On the other hand, this is not surprising since the
power of the loop accretion mechanism depends on the number and
initial sizes of the loops, and the scaling relation $\Omega_{nl} \sim \nu
\alpha$ is what should be expected from physical considerations.

For the values $\nu =1$ and $\alpha_{-2} = 1$ which are indicated by recent
cosmic string evolution simulations$^{\cite{CSsim}}$, we conclude that
the loop accretion mechanism produces enough large mass protogalaxies
to explain the observed abundance of $z \leq 4$ quasars (see Figure~1). Note
that the amplitude of the predicted protogalaxy density curves depends
sensitively on the parameters of the cosmic string scaling solution which are
still poorly determined. Hence, the important result is that there {\it are}
parameters for which the theory predicts a sufficient number of protogalaxies.
Since not all protogalaxies will actually host quasars, and since the string
parameters are still uncertain, it would be wrong to demand that the amplitude
of the $n_G$ curve agree with that of the observed $n_Q$; rather, it should lie
above the $n_Q$ curve.

It is more difficult to make definite conclusions regarding the
abundance of damped Lyman alpha absorption systems.  In the form of
Eq.~\ref{Onlres},
the condition for the cosmic string loop accretion mechanism to be
able to explain the data is also satisfied.  However, Eq.~\ref{Onlres} refers
to the value of $\Omega$ in baryonic matter.  The corresponding
constraint on the total matter collapsed in structures associated with
damped Lyman alpha systems is
$$
\Omega_{\rm DL} (z < 3) > f^{-1}_b 10^{-3}
$$
where $f_b$ is the local fraction of the mass in baryons.  From Eq.
5 it follows that the above constraint is only marginally
satisfied, and this only if the local baryon fraction $f_b$ exceeds
the average value for the whole Universe of about $f_b = 0.1$. But
in the cosmic string model with HDM we might
expect $f_b$ in nonlinear objects to be enhanced over the average
$f_b$
because after $t_{eq}$ baryons are able to cluster during the time
that
the HDM is
prevented from accreting by the free streaming. Thus, cosmic
strings may be able to
restore agreement with (~\ref{Onlres}) in a natural way.  More calculations are
required to resolve this issue.

Here we have only reported on the mechanism of forming early nonlinear
objects through accretion onto string loops. Another mechanism is
through small-scale structure on the long strings leading to the
formation of filaments rather than wakes, which has recently been
investigated$^{\cite{ZLB}}$. It was found that this could be the most effective
mechanism, and for the maximal possible amount of small-scale
structure, $\Omega_{nl} \sim 1$ can be reached already at a
redshift of $5$. It is clearly important to determine the amount
of small-scale structure present on strings.

\section{Acknowledgements}
R.M. wants to thank  Gerhard B\"orner and Deng Zugan for the invitation
to attend this workshop. We are grateful to Martin H\"ahnelt and Houjun Mo
for useful discussions.

\itemsep -0.8mm

\begin{figure}
\centerline{\psfig{figure=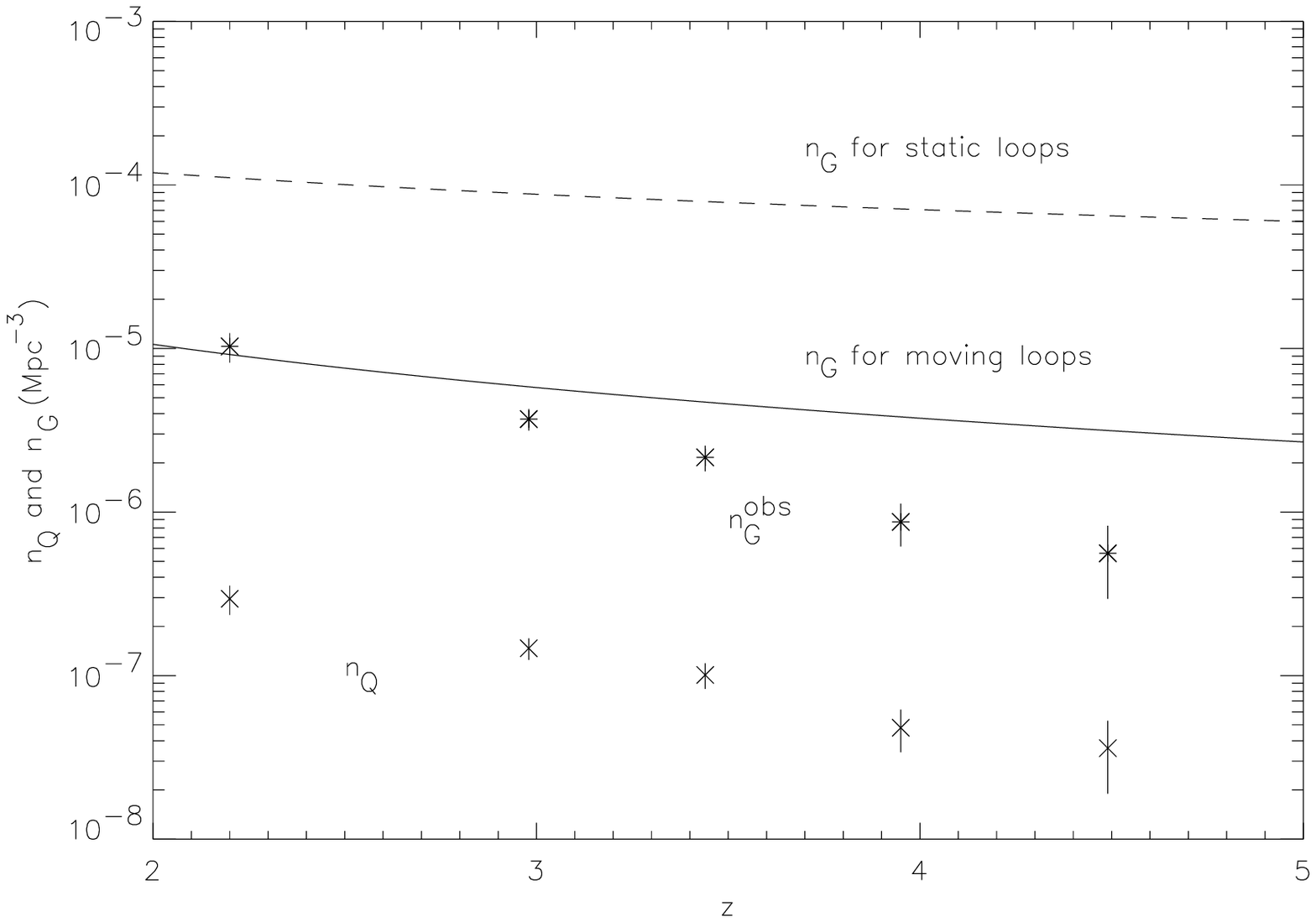}}
\caption{Comparison of the number density of
host galaxies $n_{G}^{{\rm obs}}$ (`*' marks) inferred from the
observed number density of quasars
brigher than $M_B=-26$ from the Schmidt et al. (1991,1995)
survey(`x' marks)  with the number density $n_G$ of protogalaxies of
mass greater than $10^{12} M_{\odot}$ predicted in the cosmic
string theory with HDM, for the parameters discussed in the text, and
$h_{50}=1$. The horizontal axis is the redshift.}
 \label{Fig1}
\end{figure}


\begin{thebibliography}{}
\begin{small}
\bibitem{DLAS}
{K. Lanzetta et al., {\it Ap. J. (Suppl.)} {\bf 77}, 1 (1991); \\
K. Lanzetta, D. Turnshek and A. Wolfe, {\it Ap. J. (Suppl.)} {\bf 84},
1 (1993);\\
L. Storrie-Lombardi, R. McMahon, M. Irwin and C. Hazard, `High redshift Lyman
limit and damped Lyman alpha absorbers', astro-ph/9503089, to appear in `ESO
Workshop on QSO Absorption Lines' (1995); \\
L. Storrie-Lombardi, R. McMahon, M. Irwin,{\it  MNRAS} {\bf 283}, L79 (1996).}
\bibitem{QSO}
{S. Warren, P. Hewett and P. Osmer, {\it Ap. J. (Suppl.)} {\bf
76}, 23 (1991); \\
M. Irwin, J. McMahon and S. Hazard, in `The space distribution of
quasars', ed. D. Crampton (ASP, San Francisco, 1991), p. 117;
\\
M. Schmidt, D. Schneider and J. Gunn, ibid., p. 109; \\
B. Boyle et al., ibid., p. 191; \\
M. Schmidt, D. Schneider and J. Gunn,  {\it A. J.} {\bf 110}, 68 (1995).}
\bibitem{MaB}
{C.P.. Ma, E. Bertschinger, {\it Ap. J.} {\bf 434}, L25 (1994); \\
H.J. Mo, J. Miralda-Escude, {\it Ap. J.} {\bf 430}, L25 (1994);\\
A. Klypin, S. Borgani, J. Holtzman and J. Primack, {\it Ap. J.} {\bf
444}, 1 (1995).}
\bibitem{VS 83}
{A. Vilenkin and Q. Shafi, {\it Phys. Rev. Lett.} {\bf
51}, 1716 (1983).}
\bibitem{BKST}
{R. Brandenberger, N. Kaiser, D. Schramm and N. Turok, {\it
Phys. Rev. Lett.} {\bf 59}, 2371 (1987); \\
R. Brandenberger, N. Kaiser and N. Turok, {\it Phys. Rev.} {\bf D36},
2242 (1987).}
\bibitem{MB}
{R. Moessner and R. Brandenberger, {\it MNRAS} {\bf 280}, 797 (1996),
astro-ph/9510141 .}
\bibitem{PS 94}
{K. Subramanian and T. Padmanabhan, `Constraints on the
models for structure formation from the abundance of damped Lyman
alpha systems', IUCAA preprint, astro-ph/9402006 (1994).}
\bibitem{CSrev}
{A. Vilenkin and E.P.S. Shellard, {\it 'Cosmic strings and other topological
defects'} (Cambridge Univ. Press, Cambridge, 1994); \\
M. Hindmarsh and T. Kibble, {\it Rep. Prog. Phys.} {\bf 58}, 477 (1995).; \\
R. Brandenberger, {\it Int. J. Mod. Phys.} {\bf A9}, 2117 (1994).}
\bibitem{TV86}
{T. Vachaspati, {\it Phys. Rev. Lett.} {\bf 57}, 1655 (1986).}
\bibitem{PBS}
{L. Perivolaropoulos, R. Brandenberger and A. Stebbins,
{\it Phys. Rev.} {\bf D41}, 1764 (1990); \\
R. Brandenberger, L. Perivolaropoulos and A. Stebbins, {\it Int. J. Mod. Phys.}
{\bf A5}, 1633 (1990).}
\bibitem{CMB}
{D. Bennett, F. Bouchet and A. Stebbins, {\it Ap. J. (Lett.)} {\bf 399}, L5
(1992);\\
L. Perivolaropoulos, {\it Phys. Lett.} {\bf 298B}, 305 (1993).}
\bibitem{CSsim}
{D. Bennett and F. Bouchet, {\it Phys. Rev. Lett.} {\bf 60}, 257 (1988); \\
B. Allen and E. P. S. Shellard, {\it Phys. Rev. Lett} {\bf 64}, 119 (1990); \\
A. Albrecht and N. Turok, {\it Phys. Rev.} {\bf D40}, 973 (1989).}
\bibitem{SV}
{J. Silk and A. Vilenkin, {\it Phys. Rev. Lett } {\bf 53}, 1700 (1984).}
\bibitem{AV81}
{A. Vilenkin, {\it Phys. Rev.} {\bf D23}, 852 (1981).}
\bibitem{HMM}
{T. Hara and S. Miyoshi, {\it Prog. Theor. Phys.} {\bf 81}, 1187 (1989); \\
T. Hara, S. Morioka and S. Miyoshi, {\it Prog. Theor. Phys.} {\bf 84}, 867
(1990); \\
T. Hara et al., {\it Ap. J.} {\bf 428}, 51 (1994).}
\bibitem{Vollick}
{D. Vollick, {\it Phys. Rev. D} {\bf 45}, 1884 (1992); \\
T. Vachaspati and A. Vilenkin, {\it Phys. Rev Lett.} {\bf 67}, 1057 (1991).}
\bibitem{AB95}
{A. Aguirre and R. Brandenberger, astro-ph/9505031, {\it Int. J. Mod. Phys.}
{\bf D4}, 711 (1995).}
\bibitem{Rees}
{M. Rees, {\it Mon. Not. R. Astr. Soc.} {\bf 222}, 27P (1986).}
\bibitem{Zeld}
{Ya.B. Zel'dovich, {\it Astron. Astrophys.} {\bf 5}, 84
(1970).}
\bibitem{BS89}
{R. Brandenberger and E.P.S. Shellard, {\it Phys. Rev.} {\bf D40}, 2542
(1989).}
\bibitem{EB}
{E. Bertschinger, {\it Ap. J.} {\bf 316}, 489 (1987).}
\bibitem{ZLB}
{V. Zanchin, J.A.S. Lima, R. Brandenberger, {\it Phys. Rev.} {\bf D54}, 7129
(1996), astro-ph/9607062 .}

\end{small}

\end{thebibliography}
\end{document}